\documentclass[twocolumn,english,american]{revtex4-2}
\usepackage[T1]{fontenc}
\usepackage[utf8]{luainputenc}
\usepackage{geometry}
\usepackage{amsmath}
\usepackage{amssymb}
\usepackage{graphicx}
\usepackage{babel}
\begin{document}
\title{Inhomogeneous Coulomb Models Without Defects are Sick: Domain Wall
Energies Scaling as Volume (not Boundary Area)}
\author{Garry Goldstein}
\address{garrygoldsteinwinnipeg@gmail.com}
\begin{abstract}
In this work we show that coulomb models, ones that obey a $\nabla\cdot\left\langle \mathbf{B}\right\rangle =0$
divergence free constraint - for $\left\langle \mathbf{B}\right\rangle $
being some coarse grained variables related to the microscopic degrees
of freedom of the lattice system - are sick without the inclusion
of defects with $\nabla\cdot\mathbf{B}\neq0$. We show that for generic
boundaries (ones where the fluxes of the pseudo-magnetic fields on
the boundary do not cancel: $\left(\left\langle \mathbf{B}_{R}\right\rangle -\left\langle \mathbf{B}_{L}\right\rangle \right)\cdot\mathbf{n}\neq0$
- here $\mathbf{n}$ is the unit normal and $\left\langle \mathbf{B}_{R/L}\right\rangle $
are the two ground state pseudo-magnetic fields on either side of
the domain wall) without the inclusion of defects, sharp domain walls
(on the order of the width of a unit cell) between different phases
of the system cost energy proportional to system size (not boundary
area). We present several different examples of this phenomena in
the square lattice six vertex model, in quantum dimers and in classical
spin ice in the presence of magnetic fields. We also show by example
that the condition $\left(\left\langle \mathbf{B}_{R}\right\rangle -\left\langle \mathbf{B}_{L}\right\rangle \right)\cdot\mathbf{n}=0$
is a necessary but not sufficient condition for the boundaries to
be compatible - that is domain wall energy to scale with domain wall
area and not system size. To further present the importance of boundary
conditions in Coulomb systems we show that system boundaries, even
ones that satisfy $\int_{\partial V}\mathbf{B}\cdot\mathbf{n}=0$,
have thermodynamic consequences - that is there is a cost, extensive
in system size, to the Helmholtz free energy.
\end{abstract}
\maketitle

\section{ Introduction}\label{sec:Introduction}

An interesting class of lattice systems are ones where only highly
constrained configurations are energetically allowed - those configurations
where the total ``flux'' out of any lattice site is given by zero.
For these systems, often referred to as Coulomb systems \citep{Henley_2010},
we can obtain a divergence free condition on a pseudo-magnetic field
\citep{Henley_2005,Henley_2010,Isakov_2004,Huse_2003,Moessner_2008,Youngblood_1981}.
\begin{equation}
\nabla\cdot\left\langle \mathbf{B}\right\rangle =0\label{eq:Zero-1}
\end{equation}
Furthermore Coulomb systems have many phases, liquid phases where
$\overline{\left\langle \mathbf{B}\right\rangle }=0$ and crystal
phases where generically $\overline{\left\langle \mathbf{B}\right\rangle }\neq0$.
Here $\overline{*}$ means average over configurations while $\left\langle *\right\rangle $
means coarse graining over spatial degrees of freedom to obtain smooth
functions. There are many examples of Coulomb systems - such as the
square lattice six vertex model \citep{Baxter_1982,Bleher_2014,Henley_2010,Izergin_1992,Lieb_1967,Lieb_1967(2),Lieb_1967(3),Lieb_1967(4),Lieb_1972,Franchini_2017,Lavis_1999},
quantum dimers \citep{Fisher_1961,Kasteleyn_1961,Huse_2003,Moessner_2001,Moessner_2008,Rokshar_1988,Leung_1996},
spin ice \citep{Henley_2010,Harris_1997,Ambriz_2019,Castelnovo_2008,Gringas_2009,Sakakibara_2021,Bernal_1933,Pauling_1935}
to name a few. In their liquid phases, these models admit a particularly
simple description - where the action is usually given by \citep{Henley_2010}:
\begin{equation}
Z=\int\mathcal{D}\left\langle \mathbf{B}\right\rangle \exp\left(-\frac{K}{2}\int d^{d}\mathbf{x}\left\langle \mathbf{B}\right\rangle ^{2}\left(\mathbf{x}\right)\right)\label{eq:Partition-3}
\end{equation}
Here $K$ is a model specific coupling constant. Which gives correlation
functions in $d$ dimensions: 
\begin{equation}
\overline{\left\langle \mathbf{B}_{\mu}\left(\mathbf{0}\right)\right\rangle \left\langle \mathbf{B}_{\nu}\left(\mathbf{x}\right)\right\rangle }=\frac{c_{d}}{K\left|\mathbf{x}\right|^{d}}\left(\delta_{\mu\nu}-d\hat{\mathbf{x}}_{\mu}\hat{\mathbf{x}}_{\nu}\right)\label{eq:Correlation}
\end{equation}
Where $c_{3}=4\pi$. Furthermore defects in the liquid phase which
satisfy $\nabla\cdot\left\langle \mathbf{B}\right\rangle =Q_{i}$
interact (in 3D) as:
\begin{equation}
E\left(\mathbf{x}_{1},\mathbf{x}_{2}\right)=\frac{KQ_{1}Q_{2}}{4\pi\left|\mathbf{x}_{1}-\mathbf{x}_{2}\right|}\label{eq:Coulomb_law}
\end{equation}
We note that solid phases, which are described by different parameter
regimes of the same Coulomb systems, do not admit as simple a description
as Eqs. (\ref{eq:Partition-3}), (\ref{eq:Correlation}) and (\ref{eq:Coulomb_law}).
These phases are described by a non-zero $\overline{\left\langle \mathbf{B}_{\alpha,i}\right\rangle }$
where $\alpha$ runs over the possible phases and $i$ describes the
various pseudo-magnetizations allowed for phase $\alpha$, as is allowed
by the symmetries of the model. In particular in many cases defects
in solid Coulomb phases are confined \citet{Moessner_2008}, as when
they move they leave behind a string of unfavorable configurations
which requires energy on the order of the length of the string. In
this work we explore what happens when the parameters of the Coulomb
system are inhomogeneous and different phases are preferred in different
parts of the Coulomb system. We find that generic domain walls in
Coulomb systems cost on the order of the system volume and not domain
wall area in energy. Furthermore we find that boundary conditions
for Coulomb systems (at least for the square lattice six vertex model
- which is integrable) have significant thermodynamic consequences
where specific boundaries cost on the order of the system volume in
Helmholtz free energy.

\section{ Main Idea}\label{sec:Main-Idea}

\begin{figure}
\begin{centering}
\includegraphics[width=7cm]{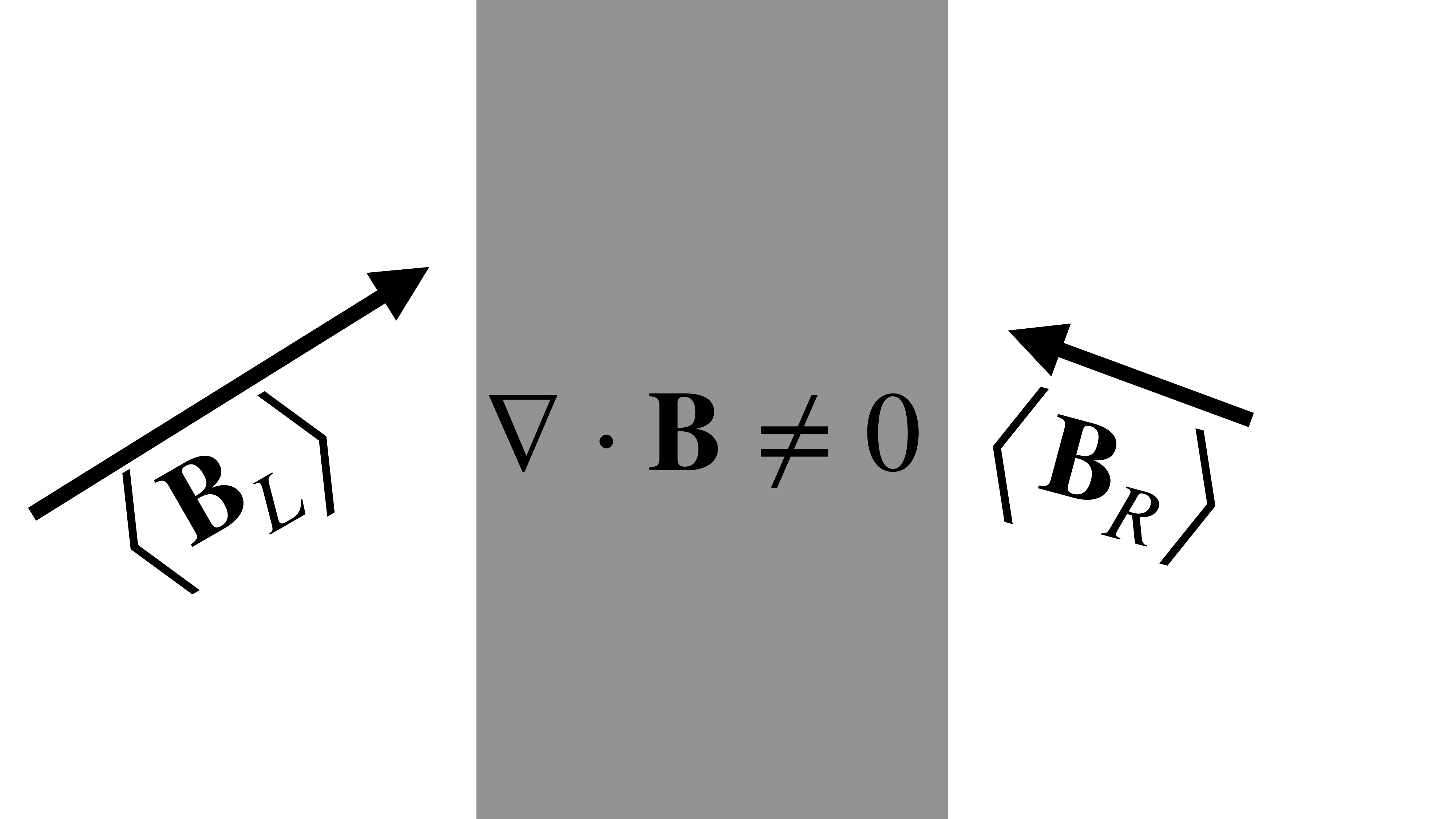}
\par\end{centering}
\caption{The main idea about non-vanishing divergence. By the divergence theorem
there are defects between different phases or there are macroscopic
changes to the system state.}\label{fig:Divergence}
\end{figure}

Consider a sharp (on the lattice constant scale) junction, where the
parameters of the model change to favor one pseudo-magnetization over
another, which furthermore satisfies: 
\begin{equation}
\overline{\left\langle \mathbf{B}_{L,i}\right\rangle }\cdot\mathbf{n}\neq\overline{\left\langle \mathbf{B}_{R,j}\right\rangle }\cdot\mathbf{n},\:\vee i,j\label{eq:Non_equal}
\end{equation}
(see Fig. \ref{fig:Divergence}). Here $L$ and $R$ represent the
magnetization of the two phases across the domain wall boundary; $i$
and $j$ run over the possible allowed $\left\langle \mathbf{B}\right\rangle $
fields for the two phases and $\mathbf{n}$ is the unit normal to
the interface (the overall sign of $\mathbf{n}$ does not matter).
Then there are three possibilities: 
\begin{enumerate}
\item There is non-zero divergence at the interface (e.g. Eq. (\ref{eq:Zero-1})
is violated by the nucleation of defects with $\nabla\cdot\mathbf{B}\neq0$).
Indeed with periodic boundary conditions by the divergence theorem
this is mandatory with $\overline{\left\langle \mathbf{B}_{L,i}\right\rangle }$
and $\overline{\left\langle \mathbf{B}_{R,j}\right\rangle }$ fixed.
In which case the domain wall costs the $\sim$ defect fugacity times
the boundary area.
\item In the case where defects are strictly forbidden with non-periodic
boundary conditions, but the $\left|\left\langle \mathbf{B}\right\rangle \right|\leq B_{M}$
(which is order one per unit cell on physical grounds for most lattice
models) the divergence can be compensated for on the boundary which
sets the boundary width to be at least: 
\begin{equation}
W\sim\frac{L\min\left|\left(\overline{\left\langle \mathbf{B}_{L,i}\right\rangle }-\overline{\left\langle \mathbf{B}_{R,j}\right\rangle }\right)\cdot\mathbf{n}\right|}{B_{M}}\label{eq:Width}
\end{equation}
(where $L$ is the linear dimensions of the system) so domain wall
has to be system size in width and the interface costs on the order
of the system volume as there is an area on the order of the system
volume where the system is not in its optimal state.
\item Alternatively in the case where defects are strictly forbidden and
the interface is not on the order of the system size the system has
to adjust so that: 
\begin{equation}
\left\langle \tilde{\mathbf{B}}_{L}\right\rangle \cdot\mathbf{n}=\left\langle \tilde{\mathbf{B}}_{R}\right\rangle \cdot\mathbf{n}\label{eq:Equal}
\end{equation}
for some $\left\langle \tilde{\mathbf{B}}_{L}\right\rangle $, $\left\langle \tilde{\mathbf{B}}_{R}\right\rangle $
not minimizing the free energy of the system, which also costs extensive
in volume energy.
\end{enumerate}
We could also have a combination of the three effects $1,2,3$. We
note that the condition in Eq. (\ref{eq:Non_equal}) is quite generically
violated for most different phases and orientations of $\mathbf{n}$.
As such Coulomb systems without defects are generically sick. Some
examples are given below. We further find, at least in the six vertex
model, that even in situations where $\int\bigtriangleup\left\langle \mathbf{B}\right\rangle \cdot\mathbf{n}=0$,
there are still pathologies where system either nucleates defects
or domain walls cost on the order of the system size.

\selectlanguage{english}%

\selectlanguage{american}%

\section{Examples}\label{sec:Examples}

We present some examples of the behavior described in the Section
\ref{sec:Main-Idea} above.

\subsection{Six vertex models on a square lattice}\label{sec:Six_vertex}

\selectlanguage{english}%
\begin{figure}
\begin{centering}
\includegraphics[width=7cm]{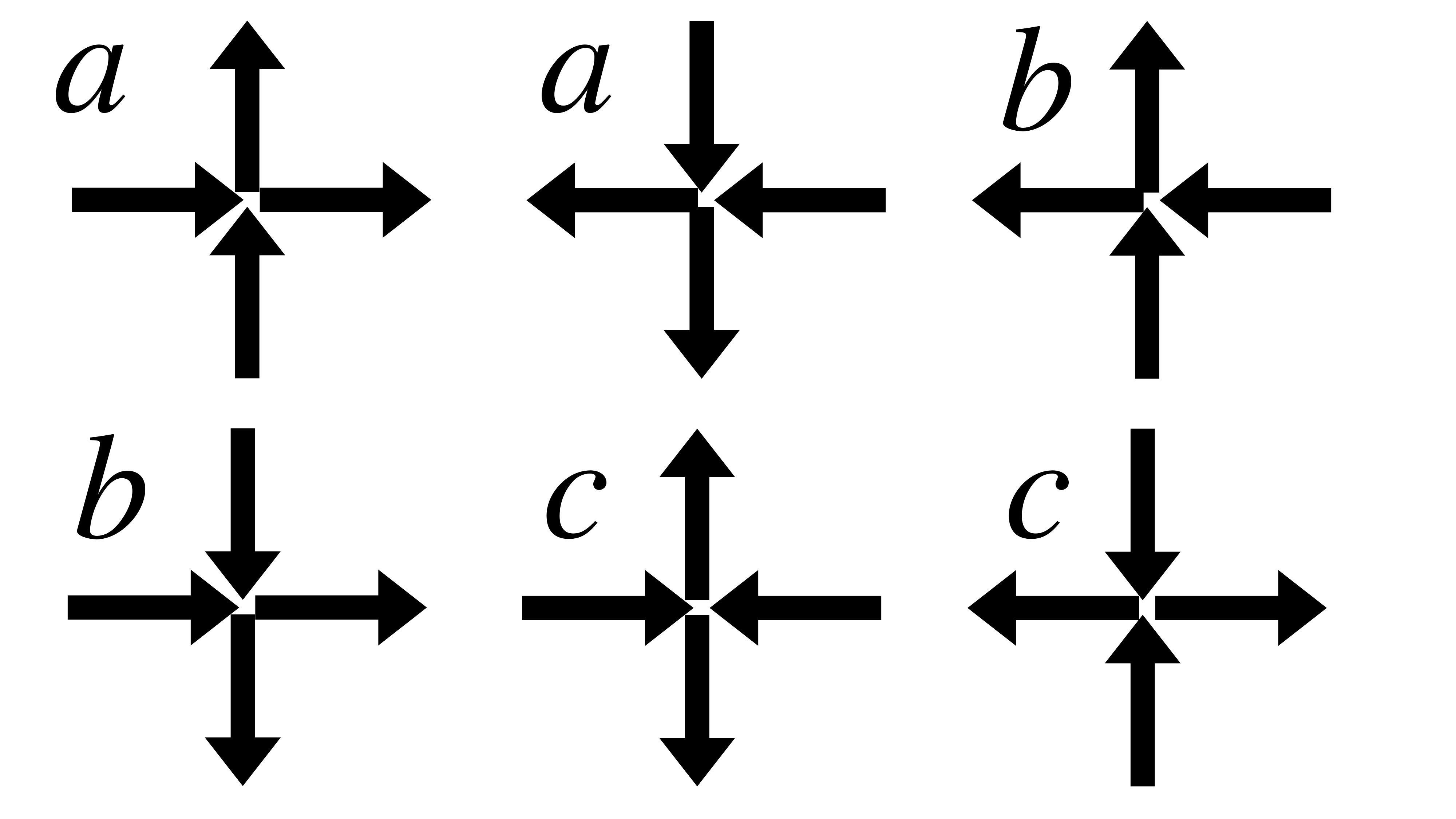}
\par\end{centering}
\caption{The six vertex configurations along with their Boltzmann weights.}\label{fig:Definitons}
\end{figure}
\foreignlanguage{american}{We consider the six vertex model \citep{Baxter_1982,Bleher_2014,Franchini_2017,Izergin_1992,Lieb_1967,Lieb_1967(2),Lieb_1967(3),Lieb_1967(4),Lieb_1972}
defined on the square lattice (see Fig. \ref{fig:Definitons}). This
model consists of spins on the links of a square lattice. There are
six possible configurations allowed, for the four links surrounding
a vertex, their Boltzmann wights are shown in Fig. \ref{fig:Definitons}.
The partition function for the six vertex model (without electric
fields) is given by:
\begin{equation}
Z=\sum_{config.}a^{N_{a}}b^{N_{b}}c^{N_{c}}.\label{eq:Partition-1}
\end{equation}
Here $N_{a/b/c}$ is the number of vertices with Boltzmann weight
$a/b/c$ in the configuration. That is a sum over all possible configurations
on the square lattice consisting of only the six types of vertices
in Fig. \ref{fig:Definitons} with a total Boltzmann weight given
by the product of the Boltzmann weights for each of the vertices.
Consider for simplicity the parameter range where:
\begin{equation}
\Delta\equiv\frac{a^{2}+b^{2}-c^{2}}{2ab}>1.\label{eq:Easy_axis}
\end{equation}
That is the easy axis (ferroelectric) limit of the six vertex model.
Then in the case $\Delta>1$ we know (see Refs. \citep{Baxter_1982,Franchini_2017})
that the model has four possible ground states shown in Fig. \ref{fig:Configurations}.
This is basically so because $c$ is so small that it does not enter
the ground state in the thermodynamic limit. The partition function
is then given by 
\begin{equation}
F=-k_{B}T\ln Z=-k_{B}TnN\max\left(\ln\left(a\right),\ln\left(b\right)\right).\label{eq:Partition}
\end{equation}
Here $n$ is the number of rows and $N$ is the number of columns.}
\begin{figure}
\begin{centering}
\includegraphics[width=7cm]{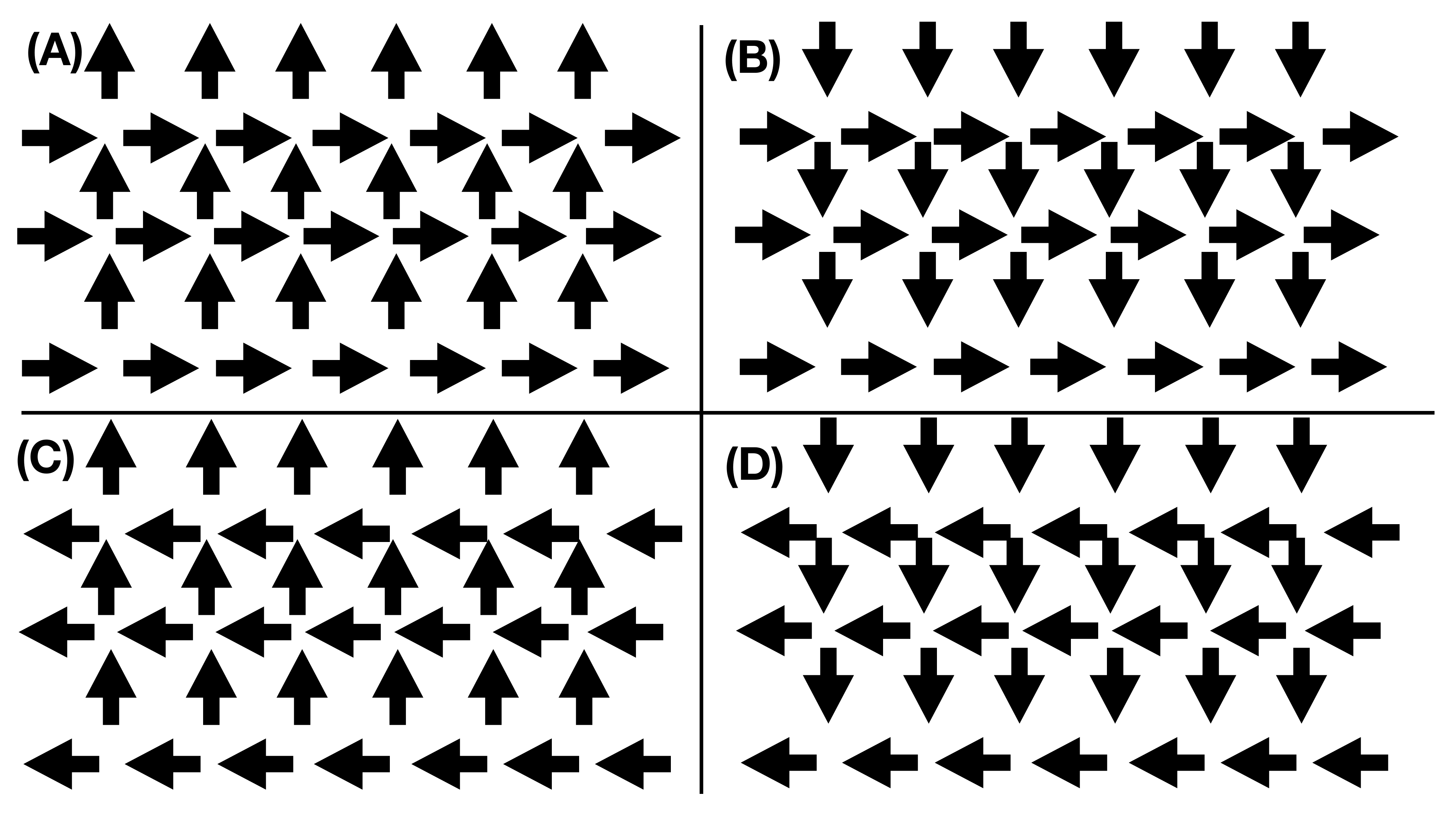}
\par\end{centering}
\caption{The four configurations most relevant to the case where $\Delta>1$.}\label{fig:Configurations}
\end{figure}

\begin{figure}
\begin{centering}
\includegraphics[width=7cm]{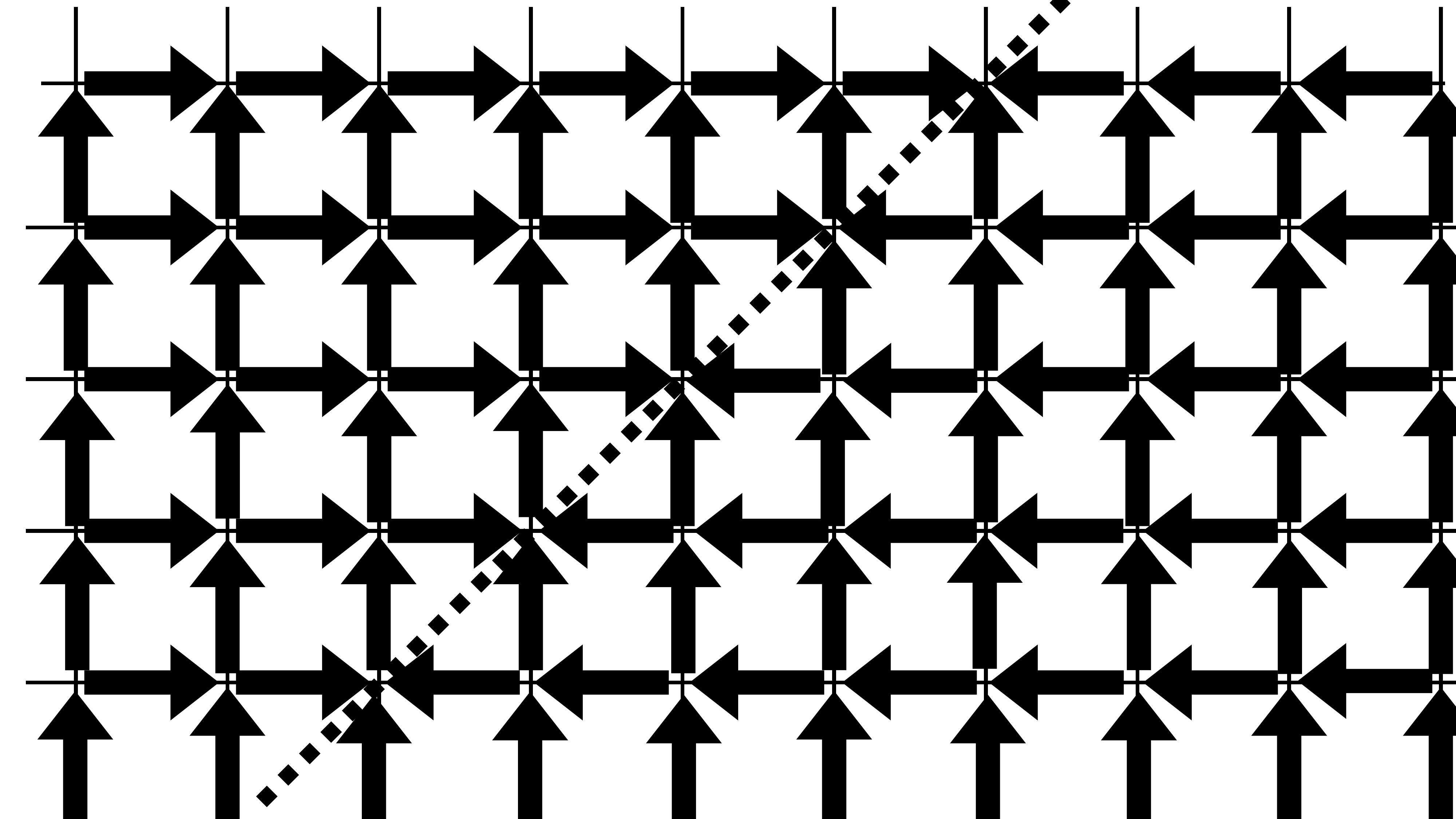}
\par\end{centering}
\caption{ Non-compatible domain wall for six vertex model.}\label{figDomain_wall_six_vertex}
\end{figure}

\selectlanguage{american}%
We introduce $\left\langle \mathbf{B}\right\rangle $ which is the
corse grained averaged spin density. Because the six allowed spin
configurations in Fig. (\ref{fig:Definitons}) have a two in - two
out structure we have that $\nabla\cdot\left\langle \mathbf{B}\right\rangle =0$.
We further note that magnetizations $\left\langle \mathbf{B}\right\rangle $
and $-\left\langle \mathbf{B}\right\rangle $ have degenerate energies.
Now consider a diagonal boundary shown in Fig. \ref{figDomain_wall_six_vertex}
then the we have that: 
\begin{equation}
\pm\left\langle \mathbf{B}_{L}\right\rangle \cdot\mathbf{n}\neq\pm\left\langle \mathbf{B}_{R}\right\rangle \cdot\mathbf{n}\label{eq:Non_equal-2}
\end{equation}
So this domain wall has a extensive in system size energy cost or
additional vertices (magnetic defects with $\nabla\cdot\left\langle \mathbf{B}\right\rangle \neq0$)
are required or there are macroscopic changes to the system state.

\selectlanguage{english}%

\selectlanguage{american}%

\subsection{Dimer models}\label{sec:Dimer_models}

\selectlanguage{english}%
\begin{figure}
\begin{centering}
\includegraphics[width=7cm]{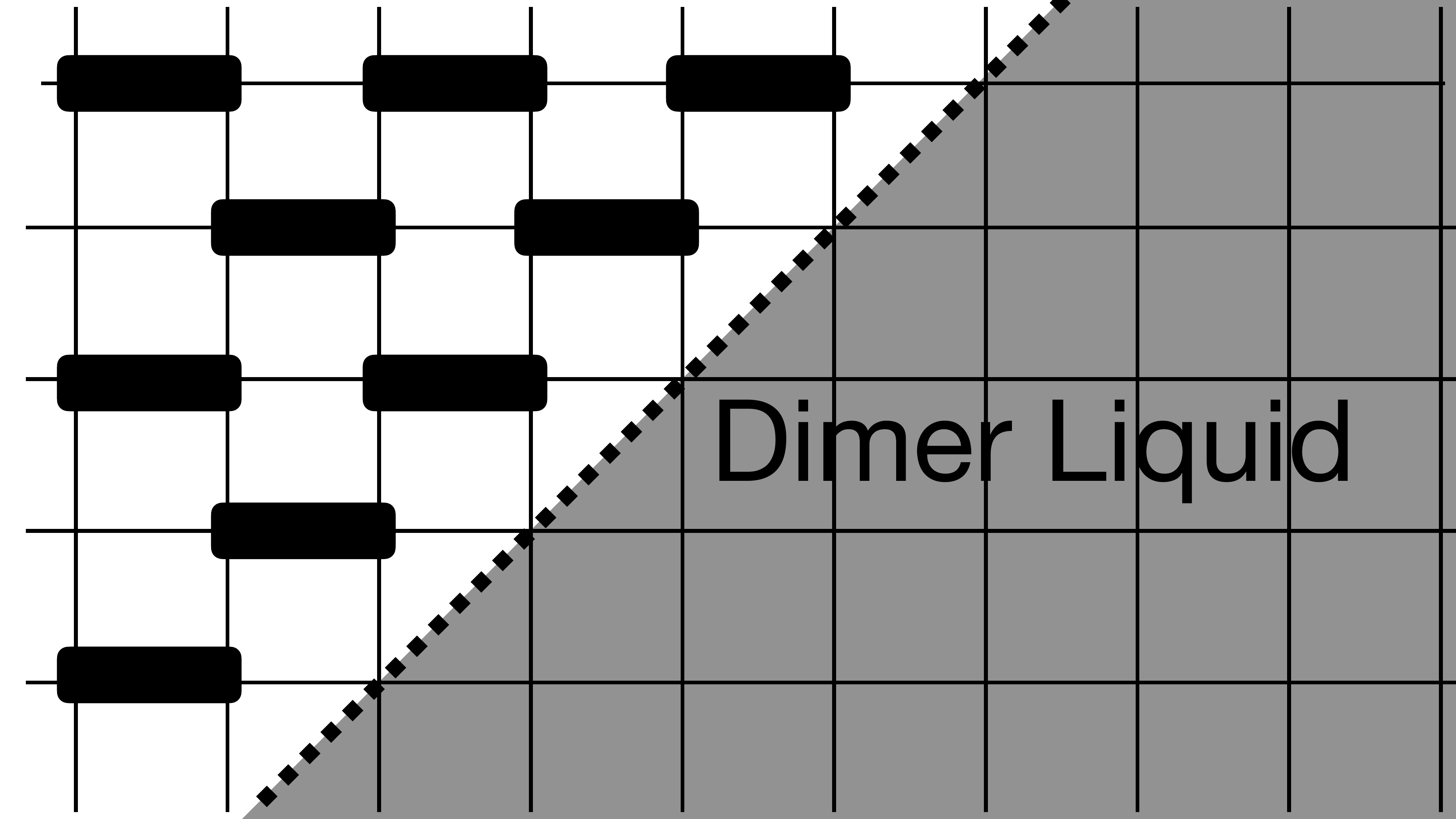}
\par\end{centering}
\caption{ Non-compatible dimer domain wall.}\label{figDomain_wall_dimer}
\end{figure}
\foreignlanguage{american}{Consider the dimer model on a square lattice
\citep{Kasteleyn_1961,Fisher_1961}. The dimers live on the links
of the lattice. For every vertex there is exactly one link attached
to the vertex with a dimer on it. The square lattice is bipartite
with two sublattices A and B. We associate a flux $z-1$ to every
link with a dimer on it - pointing between the A and B sublattices
(here $z=4$ is the co-ordination number of the lattice). For every
link without a dimer on it we assign a flux of $1$ pointing between
the B and A sublattices. We see that the flux satisfies $\nabla\cdot\left\langle \mathbf{B}\right\rangle =0$.
Now consider the quantum dimer Hamiltonian, at zero temperature, which
is given by \citep{Moessner_2008,Rokshar_1988}: 
\begin{equation}
H=-t\left(\left|\parallel\right\rangle \left\langle =\right|+\left|=\right\rangle \left\langle \parallel\right|\right)+v\left(\left|\parallel\right\rangle \left\langle \parallel\right|+\left|=\right\rangle \left\langle =\right|\right)\label{eq:Dimer}
\end{equation}
Here $t$ resonates flippable plaquettes while $v$ (depending on
its sign) penalizes or promotes flippable plaquettes. Now consider
a unit cell sharp domain wall between the liquid near the Rokhsar-Kivelson
(RK) point with $v=t$ and the magnetic phase with $v\rightarrow+\infty$
- where flippable plaquettes are highly forbidden \citep{Leung_1996,Moessner_2001,Moessner_2008}.
The dimer liquid phase has $\overline{\left\langle \mathbf{B}\right\rangle }=0$
while the magnetic solid phase has four spin configurations corresponding
to $\overline{\left\langle \mathbf{B}\right\rangle }=\pm M\left(\hat{\mathbf{x}},\hat{\mathbf{y}}\right)$
with $M\rightarrow\frac{z-2}{2}$ as $v\rightarrow+\infty$. We see
that there must be monopoles (vertices with no dimers on them) at
the interface shown in Fig. \ref{figDomain_wall_dimer} or the domain
wall boundary energy scales with the system size (not boundary length).
We note that the monopole density is $\frac{M\left(\frac{z-2}{2}\right)}{z\sqrt{2}}$
monopoles per unit length.}
\selectlanguage{american}%

\subsection{Spin Ice}\label{sec:Spin_Ice}

We consider classical spin ice (quantum effects will be neglected,
but a magnetic field will be introduced) \citep{Harris_1997,Ramirez_1999,Bramwell_2001,Bernal_1933,Pauling_1935,Castelnovo_2008,Isakov_1995,Siddharthan_1999}.
Spin ice consists of spins residing on the vertices of the pyrochlore
lattice. The pyrochlore lattice consists of corner sharing tetrahedra.
The centers of these tetrahedra form a diamond lattice - which has
two sublattices $\bar{A}$ and $\bar{B}$. We see there are two types
of tetrahedra depending on whether their centers are on the $\bar{A}$
or $\bar{B}$ sublattices. A combination of large spin orbit coupling
and strong crystal field splitting forces the spins to be Ising variables
$\sigma_{i}=\pm1$ - which lie along the line between two adjacent
tetrahedra centers on the diamond lattice. We consider the spin ice
Hamiltonian given by: 
\begin{equation}
H=J\sum_{tetra}\left(\sum_{i\in tetra}\sigma_{i}\right)^{2}\label{eq:Spin_ice}
\end{equation}
We see that there are six configurations that minimize this energy
with two spins pointing in and two spins pointing out, this leads
to a spin density with $\nabla\cdot\left\langle \mathbf{B}\right\rangle =0$
for temperatures $T\ll J$. Independent on whether the tetrahedron
center is on the $\bar{A}$ or $\bar{B}$ sublattice we can classify
the six spin ice tetrahedral configurations by $\left\{ X,\bar{X},Y,\bar{Y},Z,\bar{Z}\right\} $
where the label shows the direction of total magnetization of the
spin configuration \citep{Sakakibara_2021}. This shows the form of
the coupling of the spins to magnetic fields.

We now consider a magnetic field along the x-y plane with direction
$\mathbf{H}=h\left(\cos\left(\theta\right),\sin\left(\theta\right),0\right)$
(without loss of generality we can assume $0<\theta<\frac{\pi}{4}$),
assume that the external field terminates at a plane with basis vectors
$\left(\cos\left(\theta\right),\sin\left(\theta\right),0\right)$
and $\left(0,0,1\right)$ (we note that in this configuration $\nabla\cdot\mathbf{H}=0$
as required by Maxwell's equations). We will assume that $T\lesssim h\ll J$
so the ice-rules remain. Then we see that the states $X$ and $Y$
are dis-favored while $\bar{X}$ and $\bar{Y}$ are favored with $Z$,
$\bar{Z}$ not affected by the field. As such we now note that there
is a magnetization $\overline{\left\langle \mathbf{B}\right\rangle }=-\bar{M}\left(\cos\left(\alpha\right),\sin\left(\alpha\right),0\right)$
with $\alpha<\theta$ for tetrahedra deep in the magnetic field. This
leads to a magnetic monopole density of 
\begin{equation}
\rho=\frac{1}{2}\bar{M}\sin\left(\alpha-\theta\right)\label{eq:Density}
\end{equation}
at the location of the field termination with each monopole costing
on the order of $\sim J\sim1K$ for most spin ices.

\section{Some Counterexamples}\label{sec:Some-Counter-Examples}

Here we present some counterexamples showing that the problem presented
in Section \ref{sec:Main-Idea} is not the only problem with interfaces
and boundary conditions. We focus on the case of square lattice six
vertex model which is exactly solvable allowing us to present exact
results in our counterexamples.

\subsection{Free Fermion Line of the Square Lattice Six Vertex Model}\label{sec:Free-Fermion-Line}

\selectlanguage{english}%
\begin{figure}
\begin{centering}
\includegraphics[width=7cm]{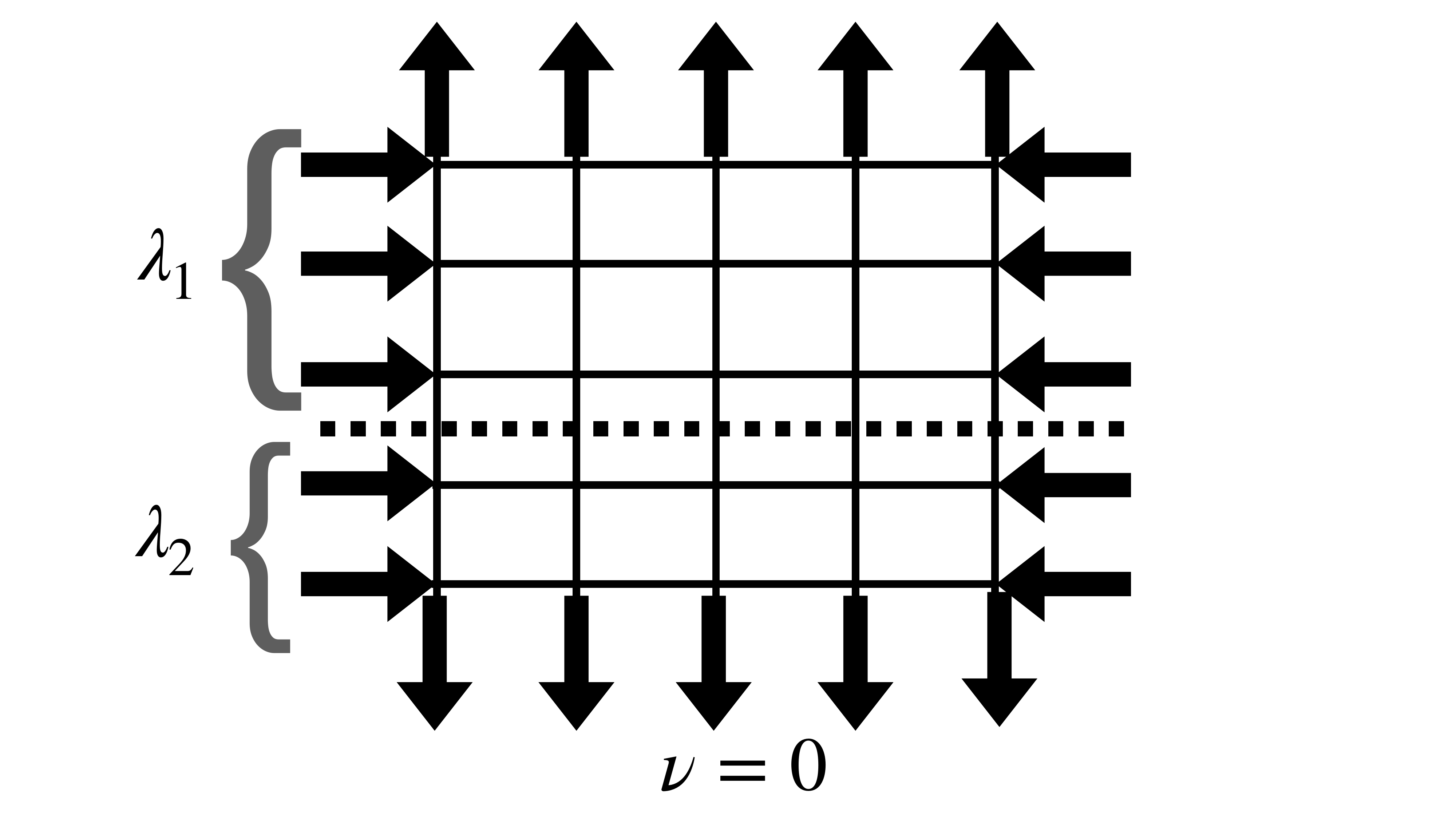}
\par\end{centering}
\caption{ The six vertex model with domain wall boundary conditions and two
rapidities.}\label{fig:Six_vertex_domain_wall}
\end{figure}

\selectlanguage{american}%
One might ask if 
\begin{equation}
\overline{\left\langle \mathbf{B}_{L,i}\right\rangle }\cdot\mathbf{n}=\overline{\left\langle \mathbf{B}_{R,j}\right\rangle }\cdot\mathbf{n},\label{eq:Equals_B}
\end{equation}
for some $i$ and $j$ is sufficient for the two sides of the domain
wall to be compatible. Here we show by example it is not. Consider
square lattice six vertex model at the free fermion point where:
\begin{equation}
a=\cos\left(\lambda_{i}-\nu_{j}\right),\:b=\sin\left(\lambda_{i}-\nu_{j}\right),\:c=1\label{eq:Free_fermion}
\end{equation}
Here $\lambda_{i}$ and $\nu_{j}$ are the inhomogeneities of the
rows and columns. Then the partition function for this system with
domain wall boundary conditions is given by \citep{Bleher_2014}:
\begin{equation}
Z=\prod_{i<j}\cos\left(\lambda_{i}-\lambda_{j}\right)\prod_{k<l}\cos\left(\nu_{k}-\nu_{l}\right).\label{eq:Partition-2}
\end{equation}
Now consider the case $\lambda_{i}=\lambda_{j}$ and $\nu_{k}=\nu_{l}$
then we have that 
\begin{equation}
Z=1\label{eq:One}
\end{equation}
Alternatively, now consider the six vertex model with domain wall
boundary conditions and two rapidities $\lambda_{1}$ and $\lambda_{2}$
with $n_{1}$ and $n_{2}$ lines respectively (see Fig. \ref{fig:Six_vertex_domain_wall})
then: 
\begin{equation}
F=-k_{B}T\ln\left(Z\right)=-k_{B}Tn_{1}n_{2}\ln\left(\cos\left(\lambda_{1}-\lambda_{2}\right)\right)\label{eq:Extensive}
\end{equation}
Which means that the additional domain wall costs extensive Helmholtz
free energy in the system volume. However both phases (with $\lambda_{1}$
and $\lambda_{2}$) are in the liquid phase with $\overline{\left\langle \mathbf{B}\right\rangle }=0$
\citep{Baxter_1982,Bleher_2014,Franchini_2017} on either side of
the interface. This makes Eq. (\ref{eq:Equals_B}) an insufficient
condition quite generically for the domain wall to have energy cost
proportional to the domain wall area and not the system volume. 

\subsection{Another Extensive in Volume Effect of Boundary Conditions}\label{sec:Another-Extensive-in}

\selectlanguage{english}%
\begin{figure}
\begin{centering}
\includegraphics[width=7cm]{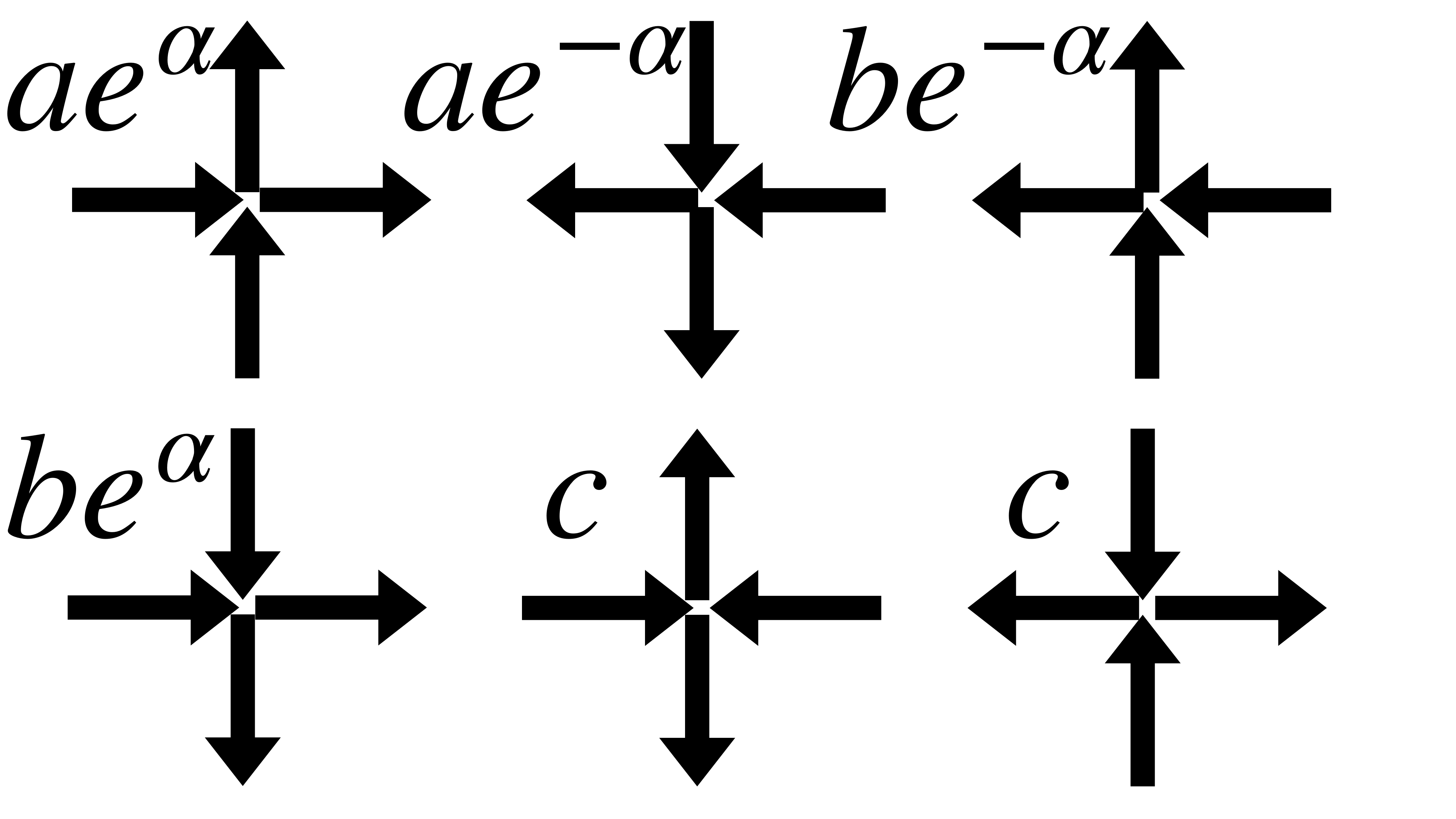}
\par\end{centering}
\caption{The six configurations along with their Boltzmann weights in an electric
field.}\label{fig:Definitons_field}
\end{figure}

\selectlanguage{american}%
One might ask if surface boundary conditions that satisfy 
\begin{equation}
\int_{\partial V}\mathbf{B}\cdot\mathbf{n}=0\label{eq:flux_free}
\end{equation}
have no extensive in system size thermodynamical consequences. Here
we show by example that this is not the case.

We will consider the six vertex model in an electric field with periodic
and domain wall boundary conditions. In which case the vertices are
given by Fig. \ref{fig:Definitons_field}. Now it is known that for
domain wall boundary conditions \citep{Bleher_2014}: 
\begin{equation}
N_{ae^{\alpha}}=N_{ae^{-\alpha}},\;N_{be^{\alpha}}=N_{be^{-\alpha}}\label{eq:Equality}
\end{equation}
(the number of $ae^{\alpha}$ and $ae^{-\alpha}$ vertices is the
same and similarly for $be^{\alpha}$ and $be^{-\alpha}$ vertices).
In particular this means that the partition function with domain wall
boundary conditions is independent of $\alpha$. Then we can point
out that it is know that with domain boundary conditions even in the
presence of a field the partition function is given by (for $\Delta>1$)
Eq. (\ref{eq:Partition}) (with $n=N$) see e.g. Ref. \citep{Bleher_2014}.
However with periodic boundary conditions the partition function (for
$\Delta>1$) is given by \citep{Baxter_1982}: $F=-k_{B}T\ln Z=$
\begin{equation}
-k_{B}TnN\max\left(\ln\left(a\right)+\alpha,\ln\left(b\right)+\alpha,\ln\left(a\right)-\alpha,\ln\left(b\right)-\alpha\right).\label{eq:Partition_field}
\end{equation}
We see that there is an extensive $k_{B}T\left|\alpha\right|$ free
energy cost to domain wall boundary conditions.

\subsection{Phase separation in the Ferroelectric Square lattice Six Vertex
Model}\label{subsec:Phase-separation-in}

\selectlanguage{english}%
\begin{figure}
\begin{centering}
\includegraphics[width=7cm]{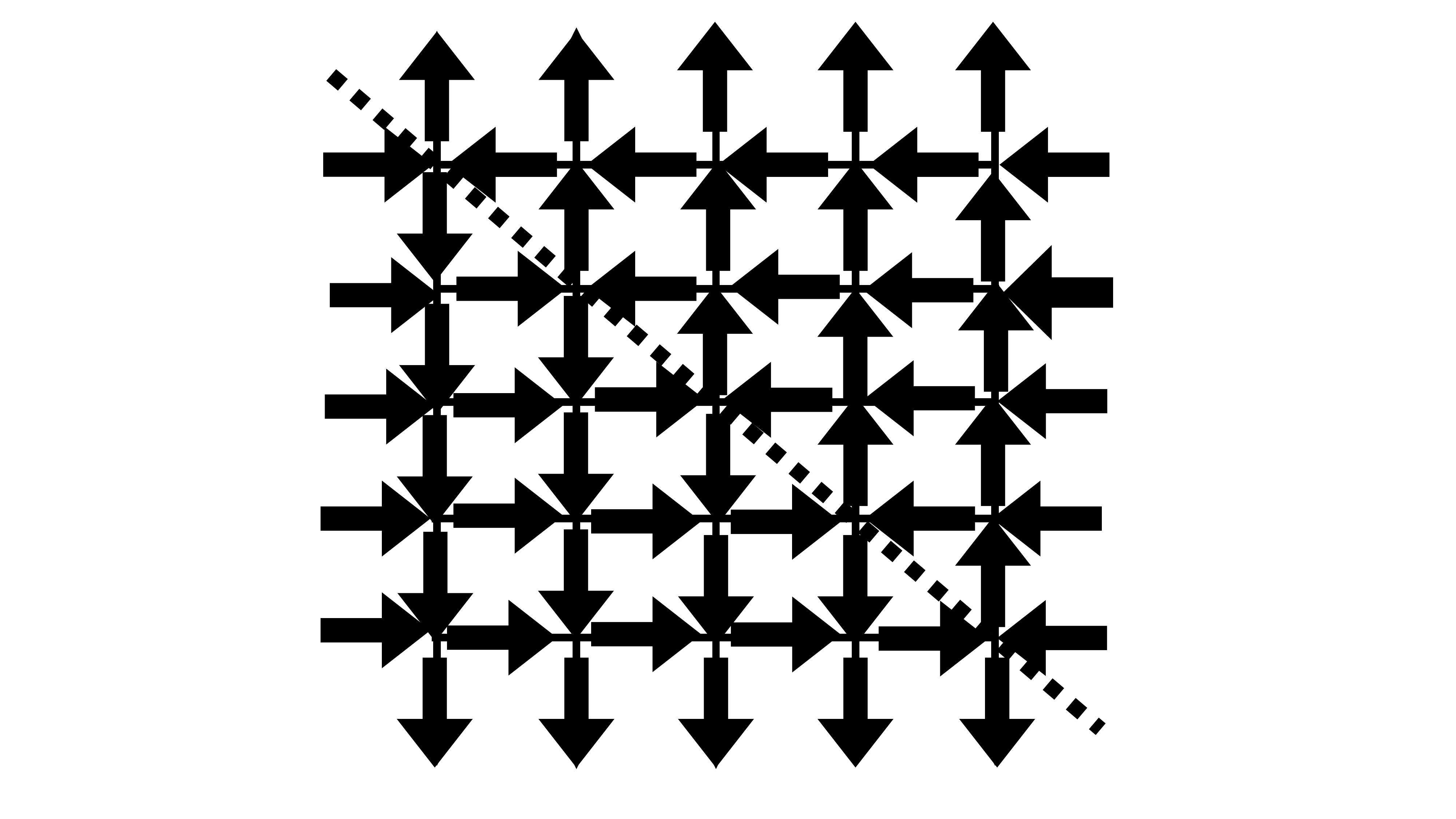}
\par\end{centering}
\caption{ The dominant contribution to the partition function of the six vertex
model with domain wall boundary conditions and $b>a+c$.}\label{fig:Phase=000020separation}
\end{figure}

\selectlanguage{american}%
One might wander if boundary conditions effect the phases of Coulomb
models. Here for the domain wall boundary conditions we show that
there is phase separation in the ferroelectric ($\Delta>1$) phase.
Indeed the dominant contribution to the partition function for $b>a+c$
is shown in Fig. \ref{fig:Phase=000020separation} (reproduced from
\citep{Bleher_2014} Fig. 8.1) where it is shown that this is the
ground state and all thermodynamically relevant states are given by
small fluctuations about the diagonal line (shown in Fig. \ref{fig:Phase=000020separation}).
We note that this corresponds to two phases with $\mathbf{B}=\pm\left(\hat{\mathbf{x}}-\hat{\mathbf{y}}\right)$
separated by a diagonal line (phase separation). On the other hand
with periodic boundary conditions there is spontaneous symmetry breaking
where where there is one phase with $\mathbf{B}=\pm\left(\hat{\mathbf{x}}-\hat{\mathbf{y}}\right)$.
This was previously observed but never considered a major effect of
boundary conditions. 

\subsection{Effect of Boundary Conditions on the Ice Point}\label{subsec:Effect-of-Boundary}

Here we further point out the importance of boundary conditions to
Coulomb systems. We consider the ice point of the six vertex model
where $a=b=c=1$. Then the partition function with periodic boundary
conditions in the thermodynamic limit is well known \citep{Baxter_1982,Lieb_1967,Lieb_1967(2),Lieb_1967(3),Lieb_1967(4),Lieb_1972}:
\begin{equation}
\ln Z_{PBC}=\frac{3}{2}N^{2}\ln\left(4/3\right)\cong0.431524N^{2}\label{eq:Parition}
\end{equation}
On the other hand the partition function with domain wall boundary
conditions is given in the thermodynamic limit by \citep{Bleher_2014}:
\begin{equation}
\ln Z_{DWBC}=N^{2}\ln\left(\frac{3\sqrt{3}}{4}\right)\cong0.261624N^{2}\label{eq:Bad}
\end{equation}
As such the partition functions at the ice point and in general the
liquid phase $-1<\Delta<1$ do not match for different boundary conditions
which leads to macroscopic effects of boundaries even without electric
fields. Indeed let us write:
\begin{align}
a & =\rho\sin\left(\lambda-t\right),\:b=\rho\sin\left(\lambda+t\right),\:c=\rho\sin\left(2\lambda\right),\nonumber \\
 & \:\rho>0,\:\left|t\right|\leq\lambda\leq\frac{\pi}{2}\label{eq:Parametrization}
\end{align}
Then let us write \citep{Baxter_1982,Bleher_2014}: 
\begin{widetext}
\begin{equation}
\ln\left(\frac{Z_{PBC}}{Z_{DWBC}}\right)=-N^{2}\left[\int_{-\infty}^{\infty}\frac{\sinh\left(2\left(\lambda+t\right)x\right)\sinh\left(\left(\pi-2\lambda\right)x\right)}{2x\sinh\left(\pi x\right)\cosh\left(\left(2\lambda x\right)\right)}+\ln\left(\frac{\pi\sin\left(\lambda+t\right)}{2\lambda\cos\left(\frac{\pi t}{2\lambda}\right)}\right)\right]\neq0\label{eq:Non_zero}
\end{equation}
\end{widetext}

Where the last statement $\neq0$ can be checked by numerical evaluation
of the integrals involved.

\section{Conclusions}\label{sec:Conclusions}

In this work we have studied inhomogeneous Coulomb systems, ones where
the model parameters depend on the position within the lattice. We
saw that in the case the model parameters sharply (on the order of
the lattice spacing) change between two different phases with different
pseudo magnetic fields $\left\langle \mathbf{B}\right\rangle $ that
satisfy Eq. (\ref{eq:Non_equal}) the Coulomb systems must either
nucleate defects to the Coulomb constraint $\nabla\cdot\left\langle \mathbf{B}\right\rangle =0$
or have domain wall energy costs that scale with the system size rather
then the domain wall length/area (this being a highly sick (pathological)
behavior). We have presented quantum dimers, spin ice and the six
vertex model as examples. We have furthermore shown that this is not
the only problem associated with Coulomb system at least in the case
of the six vertex model (which is exactly solvable). For the six vertex
model domain walls even between liquid phase $\overline{\left\langle \mathbf{B}\right\rangle }=0$
cost Helmholtz free energy proportional to system area rather then
domain wall length. Furthermore boundary conditions even ones that
satisfy Eq. (\ref{eq:flux_free}) have tremendous effects on the thermodynamics
of the system with thermodynamical changes on the order of the system
area based on boundary conditions. In the future it would be of interest
to extend these counterexamples (where not just the $\int\bigtriangleup\left\langle \mathbf{B}\right\rangle \cdot\mathbf{n}$
matters) to different non-integrable Coulomb systems such as spin
ice and quantum dimers. We would also like to study the exact behavior
of various inhomogeneous Coulomb systems with and without finite fugacity
defects in various geometries.

\textbf{Acknowledgements:} The author would like to thank Pradip Kattel
and Natan Andrei for useful discussions.

\end{document}